\newcommand{\no}{\nonumber\\}
\newcommand{\pd}{\partial}
\def\({\left(}
\def\){\right)}
\def\<{\langle}
\def\>{\rangle}
\def\tg{\tilde{G}}
\def\tgv{\tilde{G}_v}

\def\be{\begin{equation}}
\def\ee{\end{equation}}
\def\bea{\begin{eqnarray}}
\def\eea{\end{eqnarray}}

\def\d{{\delta}}
\def\D{{\Delta}}

\def\e{{\epsilon}}

\def\m{{\mu}}
\def\n{{\nu}}

\def\s{{\sigma}}

\def\ps{{\psi}}

\def\0{{(0)}}
\def\1{{(1)}}
\def\2{{(2)}}
\def\vp{{\bf p}}
\def\hp{{\hat{\bf p}}}
\def\vl{{\bf l}}
\def\vx{{\bf x}}
\def\vE{{\bf E}}
\def\vB{{\bf B}}
\def\vv{{\bf v}}
\def\vs{{\bf s}}
\def\bl{{\bar l}}
\def\bv{{\bar v}}
\def\para{\parallel}
\def\dg{\dagger}
\def\tr{\text{tr}}

\documentclass[11pt]{article}

\usepackage{amsmath}
\usepackage{amssymb}
\usepackage{bm}
\usepackage{graphicx,color}
\usepackage{slashed}
\usepackage{ulem}
\usepackage{authblk}
\usepackage{cite}

\date{\today}

\title{\bf Chiral Kinetic Theory from Effective Field Theory Revisited}

\author[1]{Shu Lin\thanks{linshu8@mail.sysu.edu.cn}}
\author[1]{Aradhya Shukla\thanks{shukla@mail.sysu.edu.cn}}
\affil[1]{School of Physics and Astronomy, Sun Yat-Sen University, Guangzhou 510275, China}

\date{\today}

\begin{document}

\maketitle

\begin{abstract}
We revisit the chiral kinetic equation from high density effective theory approach, finding a chiral kinetic equation differs from counterpart derived from field theory in high order terms in the $O(1/\m)$ expansion, but in agreement with the equation derived in on-shell effective field theory upon identification of cutoff. By using reparametrization transformation properties of the effective theory, we show that the difference in kinetic equations from two approaches are in fact expected. It is simply due to different choices of degree of freedom by effective theory and field theory. We also show that they give equivalent description of the dynamics of chiral fermions.
\end{abstract}


\newpage

\section{Introduction}
The physics of chiral fermions has attracted enormous attention recently. It has been realized that chiral anomaly can play a key role in the dynamics. The manifestation of chiral anomaly in a chiral fermion medium reveals novel transports such as chiral magnetic effect \cite{Kharzeev:2004ey,Kharzeev:2007tn,Fukushima:2008xe}, chiral separation effect \cite{Metlitski:2005pr,Son:2004tq} and chiral vortical effect etc. \cite{Erdmenger:2008rm,Banerjee:2008th,Neiman:2010zi,Landsteiner:2011cp,Landsteiner:2011iq}. There have been promising experimental signatures of these effects in systems of quark-gluon plasma \cite{Adamczyk:2014mzf,Abelev:2009ac,Abelev:2012pa,Sirunyan:2017quh} and Weyl semi-metal \cite{Li:2014bha,Gooth:2017mbd}.

In fact the above mentioned effects are the most common anomalous transports discussed in the literature. More general anomalous transports have been discussed in two complimentary frameworks. One is anomalous hydrodynamics \cite{Son:2009tf,Neiman:2010zi}, whose basic degrees of freedom are fluid velocity, local energy density and charge density. The chiral effects appears as the anomalous transports to leading order in gradient. The anomalous hydrodynamics is ideal for strongly coupled system. The opposite limit is weakly coupled system. This is the regime where the other framework chiral kinetic theory (CKT) is most suitable \cite{Son:2012bg,Son:2012wh,Son:2012zy,Stephanov:2012ki,Pu:2010as,Chen:2012ca,Hidaka:2016yjf,Manuel:2013zaa,Manuel:2014dza,Huang:2018wdl,Hattori:2019ahi,Wang:2019moi,Gao:2019znl}. Here the basic degree of freedom is distribution function of quasi-particles of fermions. It allows us to study system far from equilibrium. In CKT, chiral effects can in fact be obtained as linear response of the system to external magnetic field and vorticity field. Nonlinear response to external fields have also been obtained in CKT framework \cite{Chen:2016xtg,Gorbar:2016qfh,Hidaka:2017auj}.

The chiral kinetic equation (CKE) has been derived in different ways. They can be categorized into two approaches. One approach is field theory or equivalently Dirac equation \cite{Son:2012wh,Stephanov:2012ki,Pu:2010as,Chen:2012ca,Hidaka:2016yjf,Manuel:2013zaa,Manuel:2014dza,Huang:2018wdl,Hattori:2019ahi,Wang:2019moi,Gao:2019znl}. The other approach is effective field theory (EFT), which includes high density EFT by Son and Yamamoto (SY) \cite{Son:2012zy} and on-shell EFT by Carignano, Manuel and Torres-Rincon (CMTR) \cite{Carignano:2018gqt}. The EFT Lagrangian is defined with a cutoff. It is found by CMTR that the resulting chiral kinetic equation (CKE) differs from the counterpart obtained from field theory in higher order terms in expansion in the cutoff. We revisit the approach by SY and find agreement with CMTR.
The key to understand the difference lie in the reparametrization properties of EFT. As we shall show, while the action and equation of motion is invariant under reparametrization. The Wigner function and operators acting on it are NOT. This leads to an ambiguity in formulating CKE. We fix the ambiguity with a simple scheme. More importantly, the reparamatrization properties dictate that a difference in CKE is actually expected: the difference can be attributed to the difference in the degree of freedom chosen by the two approaches. Nevertheless, both can give correct description of dynamics of chiral fermions.

The paper is organized as follows: In Section 2, we start with high density EFT and use it to derive Wigner function and its equation of motion. Section 3 is devoted to a discussion of reparametrization invariance of the action as well as the equation of motion in EFT approach. In Section 4, we elaborate on the ambiguity in formulating CKE and present a CKE with a simple scheme. In Section 5, we show that CKE from EFT approach is equivalent to CKE from field theory despite their apparent difference. We summarize and discuss possible extension to this framework in Section 6.

\section{High Density Effective Theory and Chiral Kinetic Theory}
High density effective theory (HDET)\cite{Hong:1998tn,Hong:1999ru,Schafer:2003jn} is very useful in describing low energy dynamics. It is constructed in a simple 
manner by identifying the heavy degrees of freedom and integrating them out from the theory as irrelevant modes.
This process generates a non-local effective Lagrangian, which can be expanded in terms of large momentum. 
By construction, HDET is valid for excitations near Fermi surface. The total momentum of a particle can be decomposed as: $p^\mu = \mu  v^\mu + l^\mu$ 
with $v^\mu = (1, \bold v)$ where ${\bold v}$ is a unit velocity vector (${\bold v}^2 = 1$) denoting a point on Fermi surface. 
In this division, particle energy and momentum can be simply given by $p^0 = \mu + l^0$ and ${\bold p} = \mu {\bold v} + {\bold l}$, with 
large Fermi momentum $\mu {\bold v}$ and small residual momentum ${\bold l}$. We mention here that the choice of momentum decomposition is not unique and there is an ambiguity present in the parameter ${\bold v}$, which is connected to the reparametrization transformation in the theory. The reparametrization transformation and its implications will be discussed in a great detail in the next sections.  
The kinetic theory can be obtained from the high density effective Lagrangian by using the equations of motion for the gauge invariant Wigner function.

In the first subsection, we derive HDET Lagrangian for massless fermions valid in the vicinity of Fermi surface. This is not new but is included for completeness. Our second subsection is devoted to the derivation of Wigner function and its equation of motion, which eventually leads to dispersion relation and transport equation. 

\subsection{High Density Effective Theory}
We start with the Lagrangian for right-handed chiral fermions with finite density $\mu$ and zero temperature
\begin{eqnarray}\label{L0}
{\cal L}_{0} = \bar \psi (i \gamma^\mu D_\mu) \psi + \mu \bar\psi \gamma^0 \psi,
\end{eqnarray}
with $D_\mu = \partial_\mu + i A_\mu$ as a covariant derivative. 
Here we consider the Weyl representation for massless fermions in which, $\psi(x)$ is a two component spinor with $\gamma^\mu = (1,\,  {\boldsymbol\sigma})$. 
The energy spectrum for the above Lagrangian is given by the corresponding Hamiltonian as
\be
({\boldsymbol \sigma} \cdot {\bold p} - \mu)\psi_{\pm} = E_\pm \psi_\pm,
\ee
with  $E_\pm$ representing the energy of particles and anti-particles: $E_\pm = - \mu \pm |{\bold p}|$.
At low energy, particles near the Fermi surface with $E_+ \sim 0$ are the relevant degrees of freedom while anti-particles with $E_- \sim -2\mu$
are identified as heavy modes and integrated out from the theory.
In integrating out the heavy mode, we decompose the energy and momentum of fermions as 
$ p^0 = \mu + l^0$ and ${\bold p} = \mu {\bf v} + {\bf l}$ with $l^0, {\bf l}  \ll \mu$. The decomposition of large and small momenta
is done by taking the Fourier transform as 
\begin{eqnarray}
\psi (x) = \sum_{\bold v} e^{i \mu {\bold v} \cdot {\bf x}} \big[ \psi_{+v} (x) + \psi_{-v} (x) \big].
\end{eqnarray}
The fermion field is represented as a sum over different patches of Fermi surface, with large Fermi momentum factored out in the transformation, leaving
$\psi_{\pm v} (x)$ as the velocity dependent fields carrying the residual momentum ${\bold l}$. Further, we define the projection
operators for massless fermions: $P_\pm = (1 \pm {\boldsymbol\sigma} \cdot \vv)/2$, with the properties $P_\pm \,\psi_{\pm v}=\psi_{\pm  v}$ 
and $P_\pm \,\psi_{\mp v}=0$. They will be used to
project the positive and negative energy states $\psi_{\pm v} (x)$ from state $\psi (x)$, respectively.

For the particle near Fermi surface, Lagrangian \eqref{L0}, in terms of the new variables,  can be expressed as
\begin{eqnarray}\label{L1}
{\cal L}_{1} =  \psi_{+v}^\dagger i v \cdot D \psi_{+ v} + \psi_{-v}^\dagger (2\mu + i {\bar v} \cdot D ) \psi_{-v}
+ \psi_{+ v}^\dagger  i {\slashed D}_\perp \psi_{-v}
+ \psi_{- v}^\dagger  i {\slashed D}_\perp \psi_{+v},
\end{eqnarray}
with  ${\slashed D}_\perp = \sigma^\mu_\perp D_\mu$ with $\sigma^\mu_{\perp} = (0, {\boldsymbol \sigma}-\vv ({\bold v} \cdot {\boldsymbol \sigma}))$. In the limit $ l/\mu \rightarrow 0$, irrelevant degrees of freedom or heavy mode ($\psi_{-v}$) can be integrated out by using the classical equation of motion (EOM)
\begin{eqnarray}\label{ps-v_represent}
&&(2\mu + i {\bar v} \cdot D ) \psi_{- v} +   i {\slashed D}_\perp \psi_{+ v} = 0, \nonumber\\
&&\psi_{- v} = \frac{1}{2\mu}  \sum_n \Big(\frac{-i {\bar v} \cdot D}{2\mu} \Big)^n (- i {\slashed D}_\perp \psi_{+ v}).
\end{eqnarray}
Putting the expression of $\psi_{- v}$ in \eqref{L1} and collecting all the terms up to $O(1/\mu^2)$, we get the effective Lagrangian, 
which depends only on $\psi_{+v}$ field, as:
\begin{eqnarray}\label{EFLag}
{\cal L}_{eff} = \psi_{+ v}^\dagger \sum_n D^{(n)} \psi_{+ v}  = \psi_{+ v}^\dagger \Bigl[ i v \cdot D  + \frac{{\slashed D}^2_\perp}{2\mu}
+ \frac{{\slashed D}_\perp (-i {\bar v} \cdot D) {\slashed D}_\perp}{4\mu^2} \Bigr]\psi_{+v},
\end{eqnarray}

\subsection{Wigner Function and Equation of Motion}

We are interested in deriving the chiral kinetic theory and higher order corrections to the dispersion relation for the Weyl fermion. Thus, we construct the two-point function $G_v (x, y) = \<\psi_v (x) \psi_v^\dagger (y)\>$ corresponding to the effective field. For homogeneous system in thermal equilibrium, two-point function $G_v (x, y)$ depends only on the relative coordinates.
For inhomogeneous system, it is convenient to work with relative and central coordinates $s^\mu = x^\mu - y^\mu$ and
$X^\mu = (x^\mu + y^\mu)/2$, respectively. For the derivation of dispersion relation and transport equation, we use the Wigner function formalism \cite{Vasak:1987um,Elze:1986qd,Elze:1989un,Zhuang:1995pd}.
We define the Fourier transform of the two-point function with respect to relative variable $s^\mu$:
\be
G_v (X, l) = \int d^4s \,e^{i l \cdot s}\; G_v(x, y)\equiv \int_s \,e^{i l \cdot s}\; G_v(x, y),
\ee
where $l^\mu$ denotes the residual momentum. It is to be mentioned here that the Wigner function has similar hermiticity property 
as the two-point function but it is not gauge invariant. Thus, in the presence of gauge field, it is difficult to make the physical interpretation 
of Wigner function as a quantum analogue of distribution function. To maintain gauge invariance, Wigner transform is multiplied 
by the linking operator as follows  
\begin{eqnarray}\label{GnoU}
{\tilde G}_v (X, l) = \int_s\, e^{i l \cdot s} \;G_v (X + s/2, X - s/2) U(X -s/2, X + s/2).
\end{eqnarray}
In the above $U(y, x)$ is the Wilson line given by
\begin{eqnarray}
U (y, x) = P \,exp\, \Big[-i \int^y_x dz^\mu A_\mu (z)\Big],
\end{eqnarray}
with path ordering $P$ from $y$ to $x$.
EOM emerging from the effective Lagrangian is satisfied by the bare two-point function as
\begin{eqnarray}\label{bare_eom}
{\cal D}_x\, G_v (x, y) = 0, \qquad \quad  G_v (x, y) \,{\cal D}^\dagger_y = 0,
\end{eqnarray}
here operator ${\cal D}$ is given by ${\cal D} = D^{(0)} + D^{(1)} + D^{(2)}$. We also note that function $G_v(x,y)$ 
satisfies the properties: $P_-\, G_v (x, y) = 0, \; G_v (x, y) \,P_- = 0$, which are known as the projection conditions.
Considering the above, we can construct the following expressions by summing and subtracting the two terms in eq. \eqref{bare_eom} 
\begin{eqnarray}\label{Ipm} 
I^{(n)}_{\pm} = \int_s\, e^{i l \cdot s} \Big( {\cal D}_x \,G_v (x, y) \pm  G_v (x, y)\, {\cal D}_y^\dagger \Big).
\end{eqnarray}
We will express \eqref{Ipm} by gauge invariant Wigner function in the following. To proceed, we consider system with small inhomogeneity, the above equation can then be simplified by using the gradient expansion and rewriting the derivatives
as $\partial_x = \partial_s + \frac{1}{2} \partial_X, \, \partial_x = -\partial_s + \frac{1}{2}\, \partial_X$ with 
gauge field $A_\mu(X \pm s/2) \approx   A_\mu (X) \pm \frac{1}{2}\, (s \cdot \partial_X)\, A_\mu (X)+O(\pd_X^2)$. We perform the gradient expansion
by neglecting the terms which involves higher order derivatives $\partial_X$ to obtain  
\begin{eqnarray}\label{link}
U (y, x) =  e^{i s^\m A_\mu (X)},
\end{eqnarray}
We assume the following hierarchy of scales: $\pd_X\ll l$, spacetime disturbance is much slower than momentum so that we can ignore higher order terms in $\pd_X$; $l\ll\m$, this is needed to justify HDET, which describes low energy dynamics.
Finally, making \eqref{Ipm} gauge invariant and collecting the contributions from Wilson line, we obtain the following results:
\begin{eqnarray}\label{Ipmv}
&& I^{(0)}_+ = 2 v \cdot \bl \,{\tilde G}_v, \qquad \qquad  I^{(0)}_- = i v^\mu \D_\m {\tilde G}_v, \nonumber\\
&& I^{(1)}_+ = \frac{1}{\mu} \Big[- \bl^2_\perp +  {\bold B} \cdot {\bold v} \Big]\, {\tilde G}_v, \quad \qquad I^{(1)}_- = \frac{i}{\mu} \bl^\mu_\perp  \D_\m {\tilde G}_v, \nonumber\\
&& I^{(2)}_+ =\frac{1}{4\mu^2} \Big[4\bl_\parallel \bl^2_\perp - 4\bl_\parallel ({\bold B} \cdot {\bold v}) + 2 {\bold B} \cdot {\bold \bl_\perp} + 2({\bold E} \times {\bold \bl})\cdot {\bold v} \Big] {\tilde G}_v \nonumber\\
&&I^{(2)}_- = - \frac{i}{4\mu^2} \Big[ \Bigl( 4\bl_\parallel \bl_\perp^\mu - {\bar v}^\mu (\bl^2_\perp - {\bold B} \cdot {\bold v}) \Bigr) 
\D_\m  - \Bigl(\varepsilon^{ijk} v^k {\bar v}_\mu F^{i\mu} \Bigr) \D_j  \Big] {\tilde G}_v,
\end{eqnarray}
where we have defined $\D_\m=\pd_\m-F_{\m\n}\frac{\pd}{\pd l_\n}$. $\bl^\m=l^\m-A^\m$ is the kinetic momentum of particle. In the following, we will suppress the bar for notational simplicity. Details of the calculation is collected in appendix.
From the $I^{(n)}_+$ terms, together with the projection conditions, we deduce the following form of ${\tilde G}_v$
\be\label{Gv_para}
{\tilde G}_v = 2 \pi\,P_+\, \delta \Big( l_0 - l_\parallel - \frac{1}{2\mu} [l^2_\perp -  {\bold B} \cdot {\bold v}] + \frac{1}{2\mu^2} [l_\parallel (l^2_\perp - {\bold B} \cdot {\bold v})] + \frac{1}{4\mu^2} [{\bold B} \cdot {\bold l_\perp} + ({\bold E} \times {\bold l})\cdot {\bold v}]\Big) n_v(X,l),
\ee
where $P_+$ and $n_v(X, l)$ are the projection operators and distribution function, respectively. We point out that \eqref{Ipmv} agrees with SY \cite{Son:2012zy} up to order $O(1/\mu)$. At order $1/\mu^2$, we get different coefficients in last two terms of both $I^{(2)}_+$ and $I^{(2)}_-$, which is consistent with CMTR \cite{Carignano:2018gqt} upon identifying the cutoffs in the two effective theories. As we shall show below, the difference is crucial for understanding the kinetic equation.

The delta function in \eqref{Gv_para} is usually interpreted as dispersion relation, which naively should be invariant under reparametrization. In other words, the dispersion should not depend on $\vv$ when converting to original momentum $p^0$ and $\vp$. A quick exercise shows that this is not the case, which hints an ambiguity in the resulting CKE! We postpone writing down CKE, but investigate the reparametrization transformation more closely in the next section. The results will guide us to write down unambiguously CKE in EFT approach. In addition, it provides a resolution to the discrepancy between CKE from EFT approach and field theory approach.

\section{Reparametrization Invariance and Degree of Freedom}

Reparametrization is a redundancy in a theory which manifests that the physical implications do not change upon choosing a 
slightly different parameter. Reparametrization invariance (RI) has been discussed extensively in the heavy quark effective theory (HQET)\cite{Georgi:1990um,Chen:1993np,Kilian:1994mg,Finkemeier:1997re,Sundrum:1997ut} as well as soft collinear effective theory \cite{Bauer:2000yr,Beneke:2002ph,Manohar:2002fd,Chay:2002vy}. The symmetry greatly constrains the form of the Lagrangian for the effective theories, which is particularly useful for higher order terms. Here we closely follow the discussion of HQET, in which the field describing the particles are velocity dependent. HQET is constructed by dividing the 
particle momentum into small and large momentum part $ p^\mu = m v^\mu + l^\mu$, where $m$ and $l^\mu$ are the heavy quark mass and small 
residual momentum with $v^2 = 1$.  However, it is to be noted that this decomposition is not unique. We can have a different momentum decomposition by 
making an infinitesimal change in parameter $v^\m$ as 
\be
 v^\mu \longrightarrow  v^{\mu'} =  v^\mu + \delta  v^\mu, \qquad  l^\mu \longrightarrow  l^{\mu'} =  l^\mu - m \,\delta v^\mu,
\ee     
with a constraint $ v \cdot \delta v = 0$ emerging from the condition $v^2 = 1$. The HQET is invariant under the above reparametrization.
HDET is also reparametrization invariant. 
In HDET, we have a large chemical potential $\mu$ in place of mass parameter $m$ and the decomposition of total momentum in large 
and small parts has condition $v^2 = 0$ with $v^\mu = (1, {\bold v})$.

\subsection{Reparametrization of Classical Action}
In this subsection, we show the reparametrization invariance of the Lagrangian for massless fermions with finite density. For this purpose,
let us start with effective Lagrangian, which is non-local due to the presence of operators in the denominator
\be\label{Lag_RI}
{\cal L} = \psi^\dagger_v i v\cdot D \psi_v + \psi^\dagger_v {\slashed D}_\perp \frac{1}{2\mu + i {\bar v} \cdot D} {\slashed D}_\perp \psi_v.
\ee
This Lagrangian is essentially \eqref{EFLag}, but we keep terms to all order in $O(1/\m)$ expansion and replace $\psi_{+v}$ with $\psi_v$. It is important to mention here that the introduction of variable $v^\mu$ breaks the Lorentz invariance of the Lagrangian. 
Under the reparametrization ${\bold v} \rightarrow { \bold v'} = {\bold v} + \delta {\bold v}$, the original spinor field $\ps(x)$ does not change, but the field $\ps_v(x)$ does. The transformation of $\psi_v$ can be worked out using the following representation
\begin{align}\label{psv_represent}
\ps_v'(x)=e^{-i\m\vv'\cdot{\bf x}}P_+'\ps(x).
\end{align}
Noting that $\ps(x)$ can be expressed in terms of $\ps_v(x)$ by using classical EOM of $\ps_{-v}(x)$ \eqref{ps-v_represent}, we obtain upto $O(1/\m)$ 
\begin{align}\label{ps_represent}
\ps(x)=e^{i\m\vv\cdot\vx}\(\ps_v(x)+\ps_{-v}(x)\)=e^{i\m\vv\cdot\vx}\(1+\frac{1}{2\m}(-i{\slashed D}_\perp)\)\ps_v(x),
\end{align}
with $P_+'$ defined in terms of ${\bold v}'$ and plugging \eqref{ps_represent} into \eqref{psv_represent}, we obtain
\begin{align}\label{RT}
\psi_v \longrightarrow \psi'_v &= \psi_v + i \mu \delta  v \cdot x - \frac{ {\slashed{\delta v}}}{2} \Bigl( 1 - \frac{1}{{2\mu 
+ i {\bar v} \cdot D}} i {\slashed D}_\perp \Bigr) \psi_v, \\
\psi^\dagger_v \longrightarrow \psi^{\dagger'}_v &= \psi^\dagger - i \mu \delta  v \cdot x \psi^\dagger_v - \psi^\dagger_v  \Bigl(1 + i {\slashed D}_\perp^\dg
\frac{1}{{2\mu - i {\bar v} \cdot D^\dg}} \Bigr) \frac{ {\slashed{\delta v}}}{2} \no
&=\psi^\dagger - i \mu \delta  v \cdot x \psi^\dagger_v - \psi^\dagger_v  \Bigl(1 - i {\slashed D}_\perp
\frac{1}{{2\mu + i {\bar v} \cdot D}} \Bigr) \frac{ {\slashed{\delta v}}}{2}.\nonumber
\end{align}
Note the sign flip from $\d v\cdot x=-\d\vv\cdot{\bf x}$ with $\d v=(0,\d\vv)$. The last equality holds in the sense that integration by part is used. Each term in the transformation of $\ps_v$ can be understood as follows: the second term in \eqref{RT} arises due to a change in the Fermi momentum appearing in the Fourier decomposition. The term $1$ in the bracket follows from the change of Dirac structure in the projection operator. The last term in the bracket is reminiscent of the anti-particle contribution, because the way of integrating out anti-particle field depends on choice of $v$. We will loosely refer to this as anti-particle contribution.
We also point out that the reparametrization transformation connects terms at different orders in the $O(1/\mu)$ expansion of the Lagrangian \eqref{Lag_RI}. We will see in showing the RI, we get some mixed terms at different orders which ultimately cancel each other.

Now let us focus on RI of Lagrangian and  denote $ A = i v\cdot D +  {\slashed D}_\perp \frac{1}{2\mu + i {\bar v} \cdot D} {\slashed D}_\perp$. 
The variation of Lagrangian under reparametrization is given as
\be
\delta {\cal L} = \delta \psi^\dagger_v\, A\, \psi_v + \psi^\dagger_v\, \delta A\, \psi_v + \psi^\dagger_v\, A\, \delta\psi_v,
\ee
it is to be noted that the operator $A$ has a velocity dependence thus, its transformation is the following: 
\be
\delta A = i \delta v \cdot D + \delta {\slashed D}_\perp \frac{1}{2\mu + i {\bar v} \cdot D} {\slashed D}_\perp 
+ {\slashed D}_\perp \frac{1}{2\mu + i {\bar v} \cdot D} \delta {\slashed D}_\perp  + {\slashed D}_\perp \delta (\frac{1}{2\mu
+ i {\bar v} \cdot D} ){\slashed D}_\perp,
\ee
with $\delta {\slashed D}_\perp = - \delta v \cdot D ( {\bold v} \cdot \boldsymbol \sigma) + {\tilde v} \cdot D  {\slashed{\delta v}}$ 
and ${\tilde v}^\mu = (0,  {\bold v})$. We mention here that due to the Dirac structure and property of projection operators $P_+  \boldsymbol \sigma_\perp P_+ = 0$, 
variation $\delta A$ gives
\bea\label{RI1}
\psi_v^\dagger\,\delta A \,\psi_v &=& \psi^\dagger_v \Bigl[ i \delta v \cdot D + {\tilde v} \cdot D\frac{1}{2\mu 
+ i {\bar v} \cdot D} {\slashed {\delta v}}{\slashed D}_\perp + {\slashed D}_\perp  {\slashed {\delta v}} 
\frac{1}{2\mu + i {\bar v} \cdot D} {\tilde v} \cdot D \nonumber\\
&+& {\slashed D}_\perp \frac{1}{2\mu 
+ i {\bar v} \cdot D}   {\tilde v} \cdot D  \frac{1}{2\mu + i {\bar v} \cdot D} {\slashed D}_\perp \Bigr] \psi_v.
\eea
On the other hand, from the transformation of the other part of Lagrangian we get
\bea\label{RI2}
 \delta \psi^\dagger_v\, A\, \psi_v + \psi^\dagger_v\, A\, \delta \psi_v &=& \psi^\dagger \Bigl[ i \mu {\slashed{\delta v}} {\slashed D}_\perp \frac{1}{2\mu
 + i {\bar v} \cdot D} + i \mu \frac{1}{2\mu + i {\bar v} \cdot D} {\slashed D}_\perp {\slashed{\delta v}} \nonumber\\
&-& \frac{ {\slashed{\delta v}}}{2} {\slashed D}_\perp \frac{1}{2\mu + i {\bar v} \cdot D} {\slashed D}_\perp 
- {\slashed D}_\perp \frac{1}{2\mu + i {\bar v} \cdot D} {\slashed D}_\perp \frac{{\slashed{\delta v}}}{2} \nonumber\\
&+& i {\slashed D}_\perp \frac{{\slashed{\delta v}}}{2} \frac{1}{2\mu + i {\bar v} \cdot D} {\slashed D}_\perp i v \cdot D 
+ i v \cdot D \frac{1}{2\mu + i {\bar v} \cdot D} \frac{{\slashed{\delta v}}}{2} i {\slashed D}_\perp \nonumber\\
&+& i{\slashed D}_\perp \frac{1}{2\mu + i {\bar v} \cdot D} \frac{{\slashed{\delta v}}}{2} {\slashed D}_\perp \frac{1}{2\mu + i {\bar v} \cdot D} {\slashed D}_\perp
\nonumber\\
&+& i{\slashed D}_\perp \frac{1}{2\mu + i {\bar v} \cdot D}  {\slashed D}_\perp \frac{{\slashed{\delta v}}}{2} \frac{1}{2\mu 
+ i {\bar v} \cdot D} {\slashed D}_\perp \Bigr] \psi_v,
\eea
in the above, we have used the constraint $v\cdot\delta v = 0$.  Using  $ P_- {\boldsymbol \sigma} P_- = 0, \, \frac{\slashed{\delta v}}{2} {\slashed D}_\perp + {\slashed D}_\perp \frac{\slashed{\delta v}}{2} = - \delta v \cdot D$ from properties of Dirac structure and eqns. \eqref{RI1} and \eqref{RI2} we finally get
\be
\delta {\cal L} =  \psi^\dagger_v \big[ i \delta v \cdot D + \frac{i}{2} ( {\slashed D}_\perp {\slashed{\delta v}} 
+ {\slashed{\delta v}}{\slashed D}_\perp)\big] \psi_v = 0.
\ee 
Thus, the classical action remains invariant under the reparametrization to all order in $1/\mu$.

\subsection{Reparametrization of EOM}

In the present subsection, we concentrate on showing the RI of EOMs, from which dispersion relation, together with transport equation 
for chiral fermions, emerge from the finite density effective Lagrangian.
Following the previous reparametrization, we can write the transformation for $\psi_v, \psi_v^\dagger$ up to order $O(1/\mu)$ as follows
\bea\label{deltapsi}
&& \delta \psi_v = i\mu \delta v \cdot x\, \psi_v - \frac{\slashed {\delta v}}{2}\, \Big(1 - \frac{i}{2\mu} {\slashed D}_\perp \Big) \psi_v,  \nonumber\\
&& \delta \psi_v^\dagger = -i\mu \delta v \cdot x\, \psi_v^\dagger - \psi_v^\dagger\, \Big(1 + \frac{i}{2\mu} {\slashed D}_\perp \Big)\, \frac{\slashed {\delta v}}{2}.
\eea
It follows that the two-point function $G_v(x,y) = \<\psi_v(x) \psi_v^\dagger(y)\>$ transforms as 
\bea\label{deltaGv}
\delta G_v(x,y) &=& i\,\mu\, \delta v \cdot (x-y)\, G_v(x,y) - \frac{\slashed {\delta v}}{2}\, G_v(x, y) - G_v(x, y)\, \frac{\slashed {\delta v}}{2} \nonumber\\
&+& \frac{1}{2\mu}\, \frac{\slashed {\delta v}}{2}\, i {\slashed D}_{\perp x}\, G_v(x, y) 
- \frac{1}{2\mu}\, G_v(x, y)\, i {\slashed D}^\dagger_{\perp y}\, \frac{\slashed {\delta v}}{2}.
\eea
We are interested in the reparametrization transformation of gauge invariant Wigner function which is defined as given below
\be
\tilde{G}_v(x, y) = \int_s e^{il \cdot s}\, G_v(x, y)\, U(y, x).
\ee
It is to be noted that the gauge link $U(y, x)$ is invariant 
whereas, the residual momentum $l^\mu$ changes as $l^{\mu}{}' = l^\mu - \mu\, \delta v^\mu$. Thus, together with \eqref{deltaGv} we have
\bea
\delta {\tilde G}_v(x, y) &=& \int_s e^{il\cdot s}\;\Bigl[ \frac{1}{2\mu}\, \frac{\slashed{\delta v}}{2}\, i {\slashed D}_{\perp x} \,G_v(x, y) 
- \frac{\slashed{\delta v}}{2}\, G_v(x, y) \nonumber\\
&-& \frac{1}{2\mu}\, G_v(x, y)\, i {\slashed D}^\dagger_{\perp y}\, \frac{\slashed{\delta v}}{2}  
- G_v(x, y)\, \frac{\slashed{\delta v}}{2}  \Bigr]\, U(y, x).
\eea
Now, representing the variation $\delta {\tilde G}_v(x, y)$ by the central and relative coordinates, we  
use the gradient expansion to obtain the following
\bea\label{deltatgv}
\delta {\tilde G}_v(X, l) &=& \int_s \,\Big[- \frac{\slashed{\delta v}}{2}\, {\tilde G}_v (X, l) -  {\tilde G}_v(X, l) \,\frac{\slashed{\delta v}}{2} 
- \frac{1}{4\mu}\, \varepsilon_{jik}\, \delta v_j\, \Delta_i
\sigma^k \,{\tilde G}_v (X, l)\nonumber\\
 &+& \frac{1}{2\mu}\, \delta v_j\, l_j \,\Delta_{ij}\, {\tilde G}_v (X, l) \Big],
\eea
with definition $\Delta_{ij} = \delta_{ij} - v_i\, v_j$. According to \eqref{Gv_para}, the distribution function can be obtained by taking the trace of $\tgv(X,l)$. Note that the first two terms in \eqref{deltatgv} simply vanishes, giving rise to the following
\begin{align}\label{deltatrG}
\tr\d\tgv(X,l)=\frac{1}{4\m}\d v_j\D_iv^k\varepsilon^{ijk}\tr\tgv(X,l)+\frac{1}{2\m}\d v_jl_i\D_{ij}\tr\tgv(X,l).
\end{align}
Note that from \eqref{deltatrG}, RT of gauge invariant Wigner function comes entirely from the anti-particle contribution.

Let us focus on RI of summed and subtracted parts of equations of motion, from  which dispersion relation and transport equation emerge.
These terms are the following 
\be
I^{(n)}_{\pm} = \int_s e^{il \cdot s} ({\cal D}^{(n)}_x G_v(x, y)\pm G_v(x, y) {\cal D}^{(n)}_y ),
\ee 
where $n = 0, 1, 2,...$ For the notational simplicity, let us denote ${\cal D}_{\pm} = {\cal D}_x \pm {\cal D}^{\dagger}_y$. 
The transformation of EOM yields the following 
\bea\label{deltaEOM}
\delta \int_s\, e^{il \cdot s}\, {\cal D}_{\pm}\, G_v (x, y) &=& \int_s\, e^{il \cdot s}\, \Big[ (-i\mu \delta v \cdot s)\,{\cal D}_{\pm}\, G_v (x, y) 
+ \delta \,{\cal D}_{\pm}\, G_v (x, y) \nonumber\\
&+& {\cal D}_{\pm}\, \Bigl(i\,\mu \,\delta v \cdot s \,G_v (x, y) - \frac{\slashed {\delta v}}{2}\, G_v (x, y) 
- G_v (x, y)\,\frac{\slashed {\delta v}}{2} \nonumber\\
&+& \frac{i}{2\mu} \frac{\slashed {\delta v}}{2}\,  {\slashed D}_{\perp x}\,G_v (x, y) 
- \frac{i}{2\mu}\, G_v (x, y)\,  {\slashed D}^\dagger_{\perp y}\,\frac{\slashed {\delta v}}{2} \Bigr) \Big],
\eea
it can be easily seen that the terms in the above come from variation  of $ l^\mu, {\cal D}_{\pm}$ and $G_v(x, y)$ under reparametrization.

Let us first consider the plus EOM and use the gradient expansion for different $O({1/\mu})$ orders .
At $O(\mu)$, we have the following commutator from \eqref{deltaEOM}
\be\label{plus0}
\int_s e^{il \cdot s} [D^{(0)}_+ , i \mu \delta v \cdot s ]\, G_v (x, y) 
= \int_s\, e^{il \cdot s}\, [2\,i\, v \cdot \partial_s ,\; i \mu \delta v \cdot s ] \,G_v(X, s) = 0, 
\ee
it is very easy to observe that the above commutator vanishes due to the constraint $v \cdot \delta v = 0$. 
We have the following terms coming from $O(1)$ which cancel with each other
\be
\int_s e^{il \cdot s} \Bigl([D^{(1)}_+, i \mu \delta v \cdot s ]  + \delta D^{(0)}_+  \Bigr) G_v(X, s) = 0,
\ee
moreover, there is one more term at this order as:
\begin{align}
 \int_s e^{il \cdot s}\, {\cal D}^{(0)}_+ \big(- \frac{\slashed {\delta v}}{2}\, G_v (x, y) 
 - G_v (x, y)\, \frac{\slashed {\delta v}}{2} \big) = \int_s e^{il \cdot s}\,\Big[ 2 v \cdot l\, \Big(- \frac{\slashed {\delta v}}{2}\, G_v (X, s) 
 - G_v (X, s)\, \frac{\slashed {\delta v}}{2} \Big)\Big].
\end{align}
It vanishes upon taking the trace.
Now, at $O(1/{\mu})$, all the terms in \eqref{deltaEOM} contribute. The first three terms of \eqref{deltaEOM} are given as
 \bea
\int_s e^{il \cdot s} \Bigl (\big[D^{(2)}_+, i\mu \delta v \cdot s \big] + \delta D^{(1)}_+ \Bigr) G_v(x, y) 
&=&   \Bigl[\frac{1}{2\mu}\Big(- 4 l_\parallel l_i {\delta v}_j \Delta_{ij} 
- \varepsilon_{ijm} F_{i\n} {\delta v}_j v^m {\bar v}^\n\Big) \nonumber\\
&+& \frac{1}{\mu} \Big( l^i l^j \Big) \Big(  v_i {\delta v}_j 
+  v_j {\delta v}_i + i\varepsilon_{jkm} v_i {\delta v}_k v^m \nonumber\\
&+& i\,\varepsilon_{ikm}\, v_j\, {\delta v}_k \,v^m \Big) \Bigr] {\tilde G}_v(X, l),
 \eea
In the above, we have already taken the trace by substituting $\sigma^m \rightarrow  v^m$ and dropped the vanishing terms from the equation.
The fourth and fifth terms of \eqref{deltaEOM} arise due to change in Dirac structure, with their contribution given as: 
\bea\label{45}
\int_s e^{il \cdot s} D^{(1)}_+ \Big(- \frac{\slashed{\delta v}}{2}\, G_v (x, y) - G_v (x, y) \, \frac{\slashed{\delta v}}{2} \Big) 
&=& \frac{1}{\mu} \Big[  l_i\, l_j\, ( i \varepsilon_{jkm}\, v_i\, v_k \,\delta v^m 
- i \varepsilon_{jkm}\, v_j\, v_k\, \delta v^m)\nonumber\\
&+&   B_i\, \delta v_i \Big] \, {\tilde G}_v(X, l).
\eea
The last two terms of \eqref{deltaEOM} at $O(1/\mu)$ from the anti-particle contribution are given as follows
\bea\label{123}
\int_s e^{il \cdot s} \,D^{(0)}_+ \Bigl(\frac{1}{2\mu}\, \frac{\slashed{\delta v}}{2} \,i D_{\perp x}\,  G_v(x, y)
-   \frac{1}{2\mu} G_v(x, y)\, i D^\dagger_{\perp y}\, \frac{\slashed{\delta v}}{2}
\Bigr) &=&  \int_s e^{il \cdot s} \frac{1}{2\mu}\, i\,v \cdot \partial_s\, \Big(  - \varepsilon_{ijk} {\delta v}_i \Delta_j v^k  \nonumber\\
 &+& 2i\,{\delta v}_j \, \partial_{i s}\, \Delta_{ij}\Big)\, {\tilde G}_v (X, s),
\eea 
if we naively substitute $\partial_{s \mu} \rightarrow -i l_\mu$, we would conclude that this contribution is of higher 
order, since $v^\mu l_\mu = l_0 - l_\parallel  = O(1/\mu)$. However, we note that 
$\Delta_i = \partial_i + i s^\n F_{i\n}$, thus the term from $\pd_s$ acting on $\D_j$ should be kept:
\begin{align}\label{67}
\int_se^{il\cdot s}\Big(-\frac{1}{2\mu}\,\varepsilon_{ijk}\, {\delta v}_j\, v^m \, F_{mi}\, v^k\Big)\, {\tilde G} (X, l).
\end{align}
in the above, upon taking the trace, some of the terms vanish and we do not consider those terms.
Finally, combining all the contributions from \eqref{45}, \eqref{123} and \eqref{67}, we get
\bea
&& \frac{1}{2\mu}\, \Big( \varepsilon_{ijm}\,F_{i\n}\, \delta v_j \,v^m \,{\bar v}^\n  
- \varepsilon_{ijk}\, F_{\n i}\, \delta v_j\, v^\n \,v^k 
 + 2 B_i \,\delta v_i \Big)\,{\tilde G}_v(X, l)  \nonumber\\
&&= \frac{1}{\mu}\, \Big[\varepsilon_{min}\, \varepsilon_{ijk}\, {\delta v}_j\, v^m\, v^k\, B_n
+  B_i \, \delta v_i\,\Big]\, {\tilde G}_v(X, l) = 0.
\eea
where in the above equation, $\n = 0$  is allowed.
Thus, we have shown that the plus equations are invariant under the reparametrization transformation.

Now, let us concentrate on the minus equations. 
At order $O(\mu)$ , we have 
\be
\int_s e^{i l \cdot s}\,[D^{(0)}_-,\; i \mu \delta v\cdot s] \;G_v (x, y) 
= \int_s e^{i l \cdot s} \, [i \,v \cdot \partial_s,\; i\, \mu \,\delta v \cdot s] \;G_v (X, s)= 0,
\ee
which vanishes upon applying the constraint $v\cdot \delta v = 0,$ similar to the earlier plus equation case. 
 At $O(1)$, we have the following terms
\be
\int_s e^{i l \cdot s} \, \Bigl([D^{(1)}_-, i \,\mu \,\delta v \cdot s]  + \delta D^{(0)}_- \Bigr)\, G_v(x,y)
 = \int_s e^{i l \cdot s} (i\, \Delta_i\, \delta v_j\, \Delta_{ij} 
+ i \delta v^\mu \Delta_\mu)\, {\tilde G}_v(X, s)  = 0,
 \ee
 which cancel each other. Further, there is one more term at $O(1)$ which is: $D^{(0)}_- \big(- \frac{\slashed {\delta v}}{2} G_v (x, y) 
 - G_v (x, y) \frac{\slashed {\delta v}}{2} \big) = 0$, upon  taking the trace. 
 
 We have all terms contributing at $O(1/{\mu})$. The first three terms of \eqref{deltaEOM} are given as
 \bea\label{minus2a}
\int_s e^{il \cdot s} \Bigl (\big[D^{(2)}_-, i\mu \delta v \cdot s \big] + \delta D^{(1)}_-\Bigr) G_v(x, y) 
&=&  \frac{1}{\mu} \Big[ \frac{1}{4}\Big(- l_\n {\delta v}_j \Delta_i  - l_\n {\delta v}_i \Delta_j \nonumber\\
&-& \Delta_\n \(l_j {\delta v}_i +  l_i \delta v_j\)\Big) \Delta_{ij}\, {\bar v}^\n  
+ i\, \Delta_i\, l_j\, \Big( v_i {\delta v}_j + v_j {\delta v}_i  \nonumber\\
&+& i\varepsilon_{jkm} v_i {\delta v}^k v^m 
- i\varepsilon_{ikm} v_j {\delta v}^k v^m \Big) \Big] {\tilde G}_v (X, l),
 \eea
 in the above, trace over the sigma matrices has been taken and we have dropped vanishing terms. Moreover, we should note that
 the index $\n = 0$ is also allowed.
 The fourth and fifth terms of \eqref{deltaEOM} at this order are given as 
\be\label{minus2c}
\int_s e^{il \cdot s} D^{(1)}_- \Bigl(- \frac{\slashed{\delta v}}{2} \,G_v (x, y) - G_v (x, y)\,  \frac{\slashed{\delta v}}{2} \Bigr) 
=  \frac{1}{\mu} \Bigl[- \varepsilon_{jkm} \Delta_i  l_j v^i v^k {\delta v}^m 
- \varepsilon_{jkm} \Delta_j l_i v^i v^k {\delta v}^m \Bigr] {\tilde G}_v (X, l).
\ee
The remaining last two terms at $O(1/\mu)$ is from the anti-particle contribution, which are given as
 \bea\label{minus2b}
\int_s  e^{il \cdot s} D^{(0)}_- \Bigl(\frac{i}{2\mu}\, \frac{\slashed{\delta v}}{2}\, D_{\perp x}  G_v(x, y) 
-  \frac{i}{2\mu}\, G_v(x, y) \, D^\dagger_{\perp y} \frac{\slashed{\delta v}}{2}
\Bigr) &=&  i \,v \cdot \Delta \, \Big( -\frac{1}{4\mu}\, \varepsilon_{ijk}\, {\delta v}_i\, \Delta_j\, v^k \nonumber\\
&+& \frac{1}{\mu}\,{\delta v}_j\, l_i\, \Delta_{ij}\Big)\, {\tilde G}_v (X, l),
\eea 
the first term is ignored because $\Delta^2$ is of higher order.
Now, sum of terms \eqref{minus2a}, \eqref{minus2c} and \eqref{minus2b} vanish at $O(1/\m)$ by using $l \cdot {\bar v} = l_0 + l_\parallel =  2 l_\parallel + O (1/\mu)$. Thus, the minus equation
is also invariant under reparametrization transformation. In summary, we have shown explicitly that the action is invariant under reparametrization to all order in $1/\m$ and the EOM is invariant to order $O(1/\m)$. Viewing reparametrization invariance as a symmetry in action, we expect the EOM to be manifestly invariant to all order in $1/\m$.

Before closing this section, we elaborate on the connection of reparametrization transformation and side-jump effect \cite{Chen:2014cla,Chen:2015gta,Hidaka:2016yjf,Gao:2018wmr}. As already pointed out in CMTR, the reparametrization transformation of the Wigner function (distribution function) in fact gives rise to side-jump effect \cite{Carignano:2018gqt}. The origin of side-jump in CKT for EFT is clear from our above discussion: it comes from the dependence of effective degree of freedom on velocity $v^\m$. In case of HDET, the effective degree of freedom is particle $\ps_v$ dressed with anti-particle $\ps_{-v}$, with the dressing from integrating out the anti-particle contribution in HDET. Under variation of $v^\m$, the anti-particle contribution changes accordingly, leading to transformation of distribution function.
While formally it is similar to side-jump effect, there is a subtle difference. Our choice of velocity $v$ appears in single particle momentum decomposition and it uniquely fixes the dressed particle. It is not the same as a Lorentz boost in side-jump effect. In particular, the choice of $v$ leaves coordinate invariant in contrast to Lorentz boost. Furthermore, the Fermi sphere is not affected by choice of $v$ because the decomposition does not change the original momentum.

\section{Chiral Kinetic Theory from Effective Field Theory}

\subsection{Transport Equation}
As we show in the previous section, the Wigner function and differential operator acting on it vary separately under reparametrization. The variations cancel each other leaving the EOM invariant under reparametrization. In deriving dispersion relation and transport equation in Wigner function formalism, we use the plus equation to determine dispersion relation, and the minus equation to determine the transport equation. Now we face a puzzle: both the dispersion relation and the transport equation would be dependent on the choice of parameter. This is expected as we explained in the previous section that the degree of freedom corresponding to the Wigner function is a dressed one: particle dressed with anti-particle. The latter contribution is dependent on choice of ${\bold v}$. This is analogous to renormalization scheme dependence in field theory. There is natural choice of scheme: $\vl\para {\bold v}$, or equivalently $l_\para=l$, $l_\perp=0$. This scheme has been used in \cite{Hands:2003rw}.

Within this scheme, the plus and minus equations \eqref{Ipmv} simplify considerably as:
\begin{align}
&I_+^{\0}=2v\cdot l\tilde{G}_v, \no
&I_-^{\0}=iv^\m\D_\m\tilde{G}_v, \no
&I_+^{\1}=\frac{\vB\cdot\vv}{\m}\tilde{G}_v, \no
&I_-^{\1}=0, \no
&I_+^{\2}=-\frac{\vB\cdot\vv l}{\m^2}\tilde{G}_v, \no
&I_-^{\2}=\frac{1}{4\m^2}\big[-i\bar{v}^\m\vB\cdot\vv\D_\m+i\bar{v}^\n\e^{ijm}v^mF_{i\n}\D_j\big]\tilde{G}_v.
\end{align}
We can combine the plus equations as
\begin{align}\label{plus_para}
I_+^\0+I_+^\1+I_+^\2&=\Big[2(l_0-l)+\frac{\vB\cdot\vv}{\m}-\frac{\vB\cdot\vv l}{\m^2}\Big]\tilde{G}_v. 
\end{align}
This gives the dispersion relation $l_0=l-\frac{\vB\cdot\vv}{2\m}+\frac{\vB\cdot\vv l}{2\m^2}$. The simple scheme we choose allows us to write it in terms of original momentum $p_0=p-\frac{\vB\cdot\hp}{2p}$. This is formally the same as dispersion of particle in magnetic field. However we stress that the dispersion relation is not a physical observable. Had we chosen a different $v$, the dispersion would change accordingly.
Noting that the Wigner function satisfies $P_+\tilde{G}_v=\tilde{G}_vP_+=\tilde{G}_v$, we can parametrize $\tgv$ as
\begin{align}\label{G_nl}
\tgv=2\pi\d(l_0-l+\frac{\vB\cdot\vv}{2\m}-\frac{\vB\cdot\vv l}{2\m^2})n_v(X,\vl)P_+.
\end{align}
Here $n_v$ is the distribution function, which depends on coordinate $X$ and spatial momentum $l$. The dependence on $l_0$ is entirely in the delta function.
The transport equation follows from the minus equations. With the parametrization, it is easy to see the differential operators pass through the delta function. Thus we obtain
\begin{align}\label{minus_para}
-i(I_-^\0+I_-^\1+I_-^\2)&=\Big[\D_0+v^i(1+\frac{\vB\cdot\vv}{2\m^2})\D_i+\frac{{\bar v}^\n\e^{ijm}v^mF_{i\n}\D_j}{4\m^2}\Big] n_v(X, l) = 0.
\end{align}
The structure of transport equation is simpler if we write in terms of full momentum $\vp=\m\vv+\vl$:
\begin{align}\label{transport}
\Big[\D_0+\hp^i\(1+\frac{\vB\cdot\hp}{2p^2}\)\D_i-\frac{\e^{ijk}\hp^j\vE^k+\vB_\perp^i}{4p^2}\D_i\Big]n_v (X, l)=0.
\end{align}
We stress again the particular form holds within our scheme. It is in agreement with CMTR. However, field theory approach gives a slightly different form of transport equation \cite{Hidaka:2016yjf}:
\begin{align}\label{transport_FT}
\Big[\D_0+\hp^i\(1+\frac{\vB\cdot\hp}{2p^2}\)\D_i-\frac{\e^{ijk}\hp^j\vE^k}{2p^2}\D_i\Big]n(X,l)=0.
\end{align}
The difference in the transport equations is in fact expected. In \eqref{transport_FT}, the distribution function $n$ corresponds to particle with positive energy, while in \eqref{transport}, the distribution function $n_v$ is somewhat unconventional. It is clear from our derivation that it corresponds to an effective degree of freedom: particle dressed with anti-particle. Since the difference comes from suppressed anti-particle contribution, it is not surprising that the difference only shows up in high order terms in $1/\m$ expansion. We will show in the next section that they indeed give equivalent description of the same dynamics as expected.

\subsection{Constitutive Equation}
Let us now express physical quantities in terms of the effective distribution function $n_v$. We restrict ourselves to vector current only: $j^\m=\ps^\dg\s^\m\ps$. We wish to express it in terms of $\ps_v$. Using \eqref{ps_represent}, we obtain
\begin{align}\label{current}
j^\m=&\ps_v^\dg\(1+i{\slashed D}_\perp^\dg\frac{1}{2\m-i{\bar v}\cdot D^\dg}\)\s^\m\(1-\frac{1}{2\m+i{\bar v}\cdot D}i{\slashed D}_\perp\)\ps_v \no
=&j^{\m\0}+j^{\m\1}+j^{\m\2}+\cdots,
\end{align}
where the first three orders are given by
\begin{align}\label{j012}
j^{\m\0}&=\ps_v^\dg\s^\m\ps_v, \no
j^{\m\1}&=\frac{1}{2\m}\(\ps_v^\dg i{\slashed D}_\perp^\dg\s^\m\ps_v-\ps_v^\dg \s^\m i{\slashed D}_\perp\ps_v\), \no
j^{\m\2}&=\frac{1}{4\m^2}\(\ps_v^\dg{\slashed D}_\perp^\dg\s^\m{\slashed D}_\perp\ps_v-\ps_v^\dg{\slashed D}_\perp^\dg{\bar v}\cdot D^\dg\s^\m\ps_v-\ps_v^\dg\s^\m{\bar v}\cdot D{\slashed D}_\perp\ps_v\).
\end{align}
We proceed order by order in the evaluation of the current. At zeroth order, we simply have
\begin{align}\label{j0}
j^{\m\0}&=\ps_v^\dg\s^\m\ps_v = \tr \big[\s^\m\tgv(x,y)\big]\lvert_{x\to y}\no
&=\frac{1}{(2\pi)^4}\int_l\int_se^{il\cdot s}\tr v^\m\tgv(X,s)=\frac{1}{(2\pi)^4}\int_l\tr \big[v^\m\tgv(X,l)\big],
\end{align}
where we have made the substitution $\s^\m\to v^\m$ because $\tgv\propto P_+$. At first order, time component of the current vanishes by the trace property $\tr\s_iP_+=0$. For spatial components, we first substitute $\s$ by $\s_\perp$ by the trace property $P_+\s_\perp^i\s^j P_+= P_+\s_\perp^i\s_\perp^j P_+$:
\begin{align}
j^{i\1}&=\frac{1}{2\m}\tr\big[-i\s_\perp^i {\slashed D}_{\perp x}\ps_v(x)\ps_v(y)^\dg+i\s_\perp^i\ps_v(x)\ps_v^\dg(y){\slashed D}_{\perp y}^\dg\big]\vert_{x\to y}.
\end{align}
The limit needs to be taken carefully. We use the following identities
\bea\label{trick}
{\slashed D}_{\perp x}\ps_v(x)\ps_v(y)^\dg\vert_{x\to y} &=& \s_\perp^j D_{xj}\big[U(x,y)\tgv(x,y)\big]U(y,x)\vert_{x\to y} \nonumber\\
&=&\s_\perp^j\(\frac{1}{2}\pd_{Xj}+\pd_{sj}+\frac{i}{2}s^lF_{lj}\)\tgv(X,s)\vert_{s\to0}, \nonumber\\
\ps_v(x)\ps_v(y)^\dg{\slashed D}_{\perp y}^\dg\vert_{x\to y}&=&\big[U(x,y)\tgv(x,y)\big]\s_\perp^j D_{yj}^\dg U(y,x)\vert_{x\to y} \nonumber\\
&=&\tgv(X,s)\(\frac{1}{2}\pd_{Xj}^\dg-\pd_{sj}^\dg+\frac{i}{2}s^lF_{lj}\)\s_\perp^j\vert_{s\to0}.
\eea
Using $\s_\perp^i\s_\perp^j=\d^{ij}-v^iv^j+i\varepsilon^{ijk}v^k$, which holds in taking trace with $\tgv$, we obtain
\begin{align}\label{j1}
j^{i\1}&=\frac{1}{2\m}\tr\big[\(\D_jv^k\varepsilon^{ijk}-2i\pd_{sj}\D^{ij}\)\tgv(X,s)\big]\vert_{s\to0} \no
&=\frac{1}{2\m}\frac{1}{(2\pi)^4}\int_l\varepsilon^{ijk}\D_jv^k\tr\tgv(X,l),
\end{align}
where we have used the scheme condition $l_\perp=0$ to simplify the expression.
The second order is more complicated. For time component, only one term contributes
\begin{align}
n^{\2}&=\frac{1}{4\m^2}\ps_v^\dg{\slashed D}_\perp^\dg{\slashed D}_\perp\ps_v \no
&=\frac{1}{4\m^2}U(y,x)\tr \big[{\slashed D}_{\perp x}\tgv(x,y)U(x,y){\slashed D}_{\perp y}^\dg\big]\vert_{x\to y}.
\end{align}
We use the trick in \eqref{trick} to evaluate ${\slashed D}_{\perp x}\tgv(x,y)U(x,y){\slashed D}_{\perp y}^\dg$. Dropping $O(\pd_X^2)$ terms, we obtain
\begin{align}\label{n2}
n^{\2}&=\frac{1}{4\m^2}\tr\big[i\pd_{Xi}\pd_{sj}\varepsilon^{ijk}v^k-\pd_{si}\pd_{sj}(\D^{ij}-F_{mn}\varepsilon^{mnk}v^k)\big]\tr\tgv(X,s)\vert_{s\to0} \no
&=\frac{1}{(2\pi)^4}\int_l\frac{1}{2\m^2}\big[\vB\cdot\vv\tr\tgv(X,l)\big].
\end{align}
Spatial components of second order current contain contributions from all three terms in \eqref{j012}. The evaluation of the first term can be simplified by the identity
\begin{align}
 P_+\s_\perp^i\s^k\s_\perp^jP_+= P_+\s_\perp^iP_-\s^kP_-\s_\perp^jP_+=- P_+\s_\perp^iv^k\s_\perp^jP_+.
\end{align}
This amounts to the replacement $\s^k\to-v^k$, making the evaluation parallels the case of $n^{\2}$. It follows that
\begin{align}\label{ja2}
j_a^{i\2}=\frac{1}{(2\pi)^4}\int_l\frac{1}{2\m^2}\big[\vB\cdot\vv(-v^i)\tr\tgv(X,l)\big].
\end{align}
The other two terms can be written as
\begin{align}
j_b^{i\2}&=\frac{1}{4\m^2}U(y,x) D_{x\n}D_{xj}\tr\big[\(-\s_\perp^i\s_\perp^j{\bar v}^\n\)U(x,y)\tgv(x,y)\big]\vert_{x\to y} \no
&+\frac{1}{4\m^2}U(y,x)\tr \big[\(-\s_\perp^j\s_\perp^i{\bar v}^\n\)U(x,y)\tgv(x,y)\big]D_{yj}^\dg D_{y\n}^\dg\vert_{x\to y}.
\end{align}
After lengthy algebra, we end up with
\begin{align}\label{jb2}
j_b^{i\2}&=\frac{1}{4\m^2}\big[\(\pd_{X\n}\pd_{sj}+\pd_{Xj}\pd_{s\n}\)i\varepsilon^{ijm}v^m(-\bv^\n)+2\pd_{s\n}\pd_{sj}\D^{ij}(-{\bv}^\n)-F_{\n j}\varepsilon^{ijm}v^m(-{\bv}^\n)\big]\tr\tgv(X,s)\vert_{s\to0} \no
&=\frac{1}{(2\pi^4)}\int_l\frac{1}{4\m^2}\big[-\pd_{Xj}l_\n{\bv}^\n\varepsilon^{ijm}v^m+F_{\n j}\bv^\n\varepsilon^{ijm}v^m\big]\tr\tgv(X,l).
\end{align}
The total current is the sum of \eqref{ja2} and \eqref{jb2}:
\begin{align}\label{j2}
j^{i\2}
=\frac{1}{(2\pi)^4}\int_l\frac{1}{4\m^2}\big[-\pd_{Xj}l_\n\varepsilon^{ijm}v^m\bv^\n-2\vB\cdot\vv v^i+F_{\n j}\bv^\n v^m\varepsilon^{ijm}\big]\tr\tgv(X,l).
\end{align}
We have used the scheme condition $\vl\parallel\vv$ to simplify the expression. We have verified that our constitutive equation agrees with CMTR upon identifying cutoffs of the two effecitve theories.

\section{Equivalence of Chiral Kinetic Theories}

To show the equivalence of \eqref{transport} and \eqref{transport_FT}, we try to express $n$ in terms of $n_v$. Note that $n$ and $n_v$ are nothing but the coefficient of delta functions in $\tg$ and $\tgv$, which are defined by
\begin{align}\label{Wigners}
\tg&=\int_se^{ip\cdot s}\ps(x)\ps^\dg(y)U(y,x),\\
\tgv&=\int_se^{il\cdot s}\ps_v(x)\ps_v^\dg(y)U(y,x).
\end{align}
Using the representation of $\ps$ in terms of $\ps_v$ in \eqref{ps_represent}, we obtain
\begin{align}\label{psibarpsi}
\ps(x)\ps^\dg(y)&=e^{i\m\vv\cdot\vs}\(1-\frac{1}{2\m+i{\bar v}\cdot D_x}i{\slashed D}_{\perp x}\)\ps_v(x)\ps_v^\dg(y)\(1+i{\slashed D}^\dg_{\perp y}\frac{1}{2\m-i{\bar v}\cdot D_y^\dg}\) \no
&=e^{i\m\vv\cdot\vs}\big[\ps_v(x)\ps_v^\dg(y)+\frac{1}{2\m}\(-i{\slashed D}_{\perp x}\ps_v(x)\ps_v^\dg(y)+\ps_v(x)\ps_v^\dg(y)i{\slashed D}^\dg_{\perp y}\) \no
&+\frac{1}{4\m^2}\({\slashed D}_{\perp x}\ps_v(x)\ps_v^\dg(y){\slashed D}^\dg_{\perp y}-{\bar v}\cdot D_x{\slashed D}_{\perp x}\ps_v(x)\ps_v^\dg(y)+\ps_v(x)\ps_v^\dg(y){\slashed D}^\dg_{\perp y}{\bar v}\cdot D_y^\dg\) \no
&+O \Big(\frac{1}{\m^3} \Big).
\end{align}
Plugging \eqref{psibarpsi} into \eqref{Wigners} and taking the trace for extracting distribution function, we obtain
\begin{align}\label{GGv}
\tr\tg(X,l)=\int_se^{il\cdot s}\Big[\tr\ps_v(x)\ps_v^\dg(y)+\frac{1}{4\m^2}\tr{\slashed D}_{\perp x}\ps_v(x)\ps_v^\dg(y){\slashed D}^\dg_{\perp y}\Big]U(y,x).
\end{align}
The rest of the terms vanish by the identity $\tr\s_{\perp}^iP_+=0$. The first term in the bracket is simply $\tr\tgv(X,l)$. The second term is the higher order correction, which precisely compensate the difference in differential operator in transport equations as we shall see.

Let us evaluate the second term using the following trick
\begin{align}\label{correction}
&\int_se^{il\cdot s}\tr{\slashed D}_{\perp x}\ps_v(x)\ps_v^\dg(y){\slashed D}^\dg_{\perp y}U(y,x) \no
=&\int_se^{il\cdot s}\tr{\slashed D}_{\perp x}\tgv(X,s)U(x,y){\slashed D}^\dg_{\perp y}U(y,x) \no
=&\int_se^{il\cdot s}\tr\(\frac{1}{2}\D_i+\pd_{si}\)\s_\perp^i\tgv(X,s)\s_\perp^j\(\frac{1}{2}\D_j^\dg-\pd_{sj}^\dg\) \no
=&\int_se^{il\cdot s}\(\frac{1}{2}\D_i-il_i\)\tgv(X,s)\(\frac{1}{2}\D_j^\dg+il_j\)\tr\s_\perp^iP_+\s_\perp^j.
\end{align}
The trace can be evaluated as
\begin{align}
\tr\s_\perp^iP_+\s_\perp^j=\tr P_+\(\d_{ij}-v^iv^j+i\varepsilon^{ijk}v^k\).
\end{align}
Plugging this into \eqref{correction}, we find the symmetric terms are either $O(\D^2)$ thus are neglected or $O(l_\perp^2)$, which vanishes by our scheme condition. Keeping the anti-symmetric term, we end up with
\begin{align}\label{nnv}
\tr\tg=\tr\tgv-\frac{1}{4\m^2} l_i\D_j\tr\tgv\varepsilon^{ijm}v^m\quad\Rightarrow\quad n=n_v-\frac{1}{4\m^2}l_i\D_jn_v\varepsilon^{ijm}v^m.
\end{align}
Plugging \eqref{nnv} into \eqref{transport_FT}, we find the correction give rise to two additional terms upto $O(\frac{1}{\m^2})$, which are from $\D_0+v^i\D_i$ acting on the correction. To see them more explicitly, we expand
\begin{align}
\D_0+v^i\D_i=\pd_0-E^i\frac{\pd}{\pd l_i}+v^i\(\pd_i+\varepsilon^{ijk}B^k\frac{\pd}{\pd l_j}\).
\end{align}
The $l$-derivative terms give rise to
\begin{align}
&-\frac{1}{4\m^2}\(-E^i\frac{\pd}{\pd l_i}+v^i\varepsilon^{ijm}B^m\frac{\pd}{\pd l_j}\)l_k\D_ln_v\varepsilon^{kln}v^n \no
=&\frac{1}{4\m^2}\(E^i\D_jv^k\varepsilon^{ijk}-B_\perp^i\D_i\)n_v.
\end{align}
The generated terms precisely match the difference between \eqref{transport} and \eqref{transport_FT}. Therefore we have shown the equivalence of CKE from field theory and CKE from EFT within our simple scheme. Combining with the reparametruization invariance of EOM, which is just the transformation of CKE under change of scheme, we can conclude that the equivalence holds for arbitrary schemes as well.

\section{Summary}

We revisit the high density effective theory approach to CKT. We find the resulting CKE differs from the counterpart from field theory approach in high order terms in the $1/\mu$ expansion. Our CKE from high density effective theory is formally the same as the counterpart obtained from on-shell effective field theory upon identifying the expansion parameters in the two theories. We further show that despite different forms of kinetic equations obtained from field theory approach and effective theory approach, the two equations are equivalent, with the difference being in the choice of degree of freedoms. CKE from field theory uses particle as degree of freedom, while CKE from effective field theory uses dressed particle as degree of freedom, which follows from integrating out anti-particle contribution.

The way of integrating out the anti-particle contribution depends on the choice of a parameter in the EFT. In high density effective theory, this parameter is the Fermi velocity $\vv$. Both distribution function and CKE transform under reparametrization of $v$. Making a specific choice of $\vv\para\vl$ leads to our CKE. Similar reparametrization transformation also exists in on-shell effective theory. The transformation of distribution function upon change of $v$ is formally the same as the side-jump effect, which is the transformation of distribution function under Lorentz boost. However, there is a subtle difference between the two: unlike Lorentz boost, the reparametrization of $v$ affects neither the coordinate nor the Fermi sphere. It would be interesting to explore further possible connection with side jump. We leave it for future work.

It is worth noting that our current study does not include a collision term for fermions, therefore we do not have a mechanism for relaxation. The collision term for fermions can be included by making external gauge field dynamical. This would also introduce gauge field degree of freedom into the CKE.

Finally it is interesting to speculate possible extensions with effective theory approach to CKT. Field theory approach essentially assumes an $\hbar$ expansion in deriving CKE. It is complicated to go to higher order term in $\hbar$ expansion \cite{Gao:2018wmr}. Effective field theory assumes a different expansion in the cutoff of the EFT. In the absence of collision term, the CKE we obtain also stops at order $O(\hbar)$. In principle it is possible to go to higher order term in $\hbar$ with effective field theory. It would be interesting to further investigate this point.

\section{Acknowledgments}
We are grateful to Y. Hidaka, J-.H. Gao, R. Pisarski and Y. Yin for useful discussions. We also thank the University of Science and Technology of China for providing a stimulating environment in the ATHIC meeting 2018, during which part of this work is done. This work is supported by NSFC under Grant Nos 11675274 and 11735007 and One Thousand Talent Program for Young Scholars.

\appendix
\vskip 0.8cm
\noindent
\section{Evaluation of $I^{(n)}_{\pm}$ (with $n = 0, 1, 2$) using the Gradient Expansion}
In the appendix, we explicitly show the derivation of $I^{(n)}_\pm$ terms at different orders of $1/\mu$. As we mentioned before, form 
of the terms $I^{(n)}_\pm$ is following
\be\label{ap1}
I^{(n)}_{\pm} = \int_s\, e^{il \cdot s}\, \Big( {\cal D}^{(n)}_x\, G_v (x, y) \pm G_v (x, y)\, {\cal D}^{(n)\dagger}_y  \Big),
\ee
with ${\cal D}^{(n)} = D^{(0)} + D^{(1)} + D^{(2)}$. Considering the central and relative coordinates ($X^\mu,\, s^\mu$), with definition of 
derivative and gauge field: $\partial_x =  \frac{1}{2} \partial_X + \partial_s, \,\partial_y =  \frac{1}{2} \partial_X - \partial_s $ 
and $A_\mu (X\pm\frac{s}{2}) = A_\mu (X) \pm \frac{1}{2} s \cdot \partial_X A_\mu (X)$, we write covariant derivatives as
\bea\label{ap2}
D_\mu (x) = \frac{1}{2} \partial_{X \mu} + \partial_{s \mu} + i A_\mu (X) + \frac{i}{2} s \cdot \partial_X A_\mu (X),\nonumber\\
D^\dagger_\mu(y) = \frac{1}{2} \partial_{X \mu}^\dg - \partial_{s \mu}^\dg - i A_\mu (X) + \frac{i}{2} s \cdot \partial_X A_\mu (X).
\eea
Using the above form of covariant derivatives and expressing $G_v(x,y)=U(x,y)\tgv(x,y)=e^{-is\cdot A}\tgv(x,y)$, we can write terms in $I_\pm^{(0)}$ as
\bea\label{ap3}
iv^\mu D_{x\mu }\, \Big( {\tilde G }\, e^{-is \cdot A} \Big) &=& iv^\mu\, \Big[\frac{1}{2}\, \partial_{X \mu} + \partial_{s \mu} + i\, A_\mu 
 + \frac{i}{2} s \cdot \partial_X A_\mu \Big]\, \Big({ e^{-is \cdot A} \tilde G}_v \Big)  \nonumber\\
&=& e^{-is \cdot A}\, \Big( \frac{1}{2} \partial_{\mu x}
 + \partial_{\mu s} + \frac{1}{2} s^\nu F_{\mu\nu} \Big)\, {\tilde G}_v, \nonumber\\
 i \Big( {\tilde G }\, e^{-is \cdot A} \Big)\,  D^\dagger_{y \mu } v^\mu &=& iv^\mu\, \Big[\frac{1}{2} \partial_{X \mu} - \partial_{s \mu} - i\, A_\mu 
 + \frac{i}{2} s \cdot \partial_X A_\mu \Big]\, \Big( e^{-is \cdot A}\, {\tilde G}_v \Big) \nonumber\\
 &=& e^{-is \cdot A}\, \Big( \frac{1}{2} \partial_{\mu x} 
- \partial_{\mu s} + \frac{1}{2} s^\nu F_{\mu\nu} \Big)\, {\tilde G}_v,
\eea
sum of above terms gives us
\bea \label{ap4}
I^{(0)}_+ &=& \int_s  e^{i (l - A) \cdot s}\, 2\,i\, v^\mu \,\partial_{\mu s}\, {\tilde G}_v(x,y)  
= \int_s e^{i \bl \cdot s}\, 2\,i\, v^\mu\, (-i\, \bl_\mu)\, {\tilde G}_v(x,y) \nonumber\\
 &=&  2\, (\bl_0 - \bl_\| ) \,{\tilde G}_v(X, l),
\eea
in the above, we have used kinetic momentum $\bl_\mu=l_\m-A_\m$. 
Similarly, difference of the two terms gives
\be\label{ap5}
\int_s \, e^{i (l - A) \cdot s} \,i\,\Big(\partial_{\mu X} + i\, s^\nu F_{\nu\mu} \Big)\, {\tilde G}_v(x,y)
= i \,v^\mu \,\Big( \partial_{X \mu} - F_{\mu\nu} \frac{\partial}{\partial l_\nu} \Big)\, {\tilde G}_v (X, l).
\ee
At the order $O(1/\m)$, we have $D^{(1)} = \frac{{\slashed D}^2}{2\mu},$ from which $D^{(1)}_x = \frac{1}{2\mu}\,D_{x i}\, D_{x j} \,\sigma_{\perp i}\, \sigma_{\perp j}$ and $D^{(1) \dagger}_y = \frac{1}{2\mu}\, D^{\dagger}_{y i} \,D^{\dagger}_{y j}\, \sigma_{\perp i}\, \sigma_{\perp j}$ lead to
\bea\label{ap6}
D_{x i} \,D_{x j}\, \Big( e^{i s \cdot A} \,{\tilde G}_v\Big) &=&  e^{i s \cdot A}\, \Big[ \frac{1}{4}\, \partial_{X i}\, \partial_{X j} 
+ \frac{1}{2}\, \partial_{X i}\, \partial_{s j} + \frac{1}{2}\, \partial_{s i}\, \partial_{X j} + \partial_{s i} \partial_{s j} 
+ \frac{i}{2} F_{ij} \nonumber\\
&+& \frac{i}{2} s^m F_{mi} (\frac{1}{2} \partial_{X j} + \partial_{s j}) + \frac{i}{2}\, s^m F_{mi} \Big(\frac{1}{2}\, \partial_{X i} 
+ \partial_{s i} \Big) \Big]\, {\tilde G}_v, \nonumber\\
D^\dagger_{y j} \,D^\dagger_{y i}\, \Big( e^{i s \cdot A} \,{\tilde G}_v\Big) &=&  e^{i s \cdot A} \Big[ \frac{1}{4} \partial_{X i} \partial_{X j} 
- \frac{1}{2} \partial_{X i} \partial_{s j} - \frac{1}{2}\, \partial_{s i} \,\partial_{X j} + \partial_{s i}\, \partial_{s j} 
+ \frac{i}{2} F_{ij} \nonumber\\
&+& \frac{i}{2}\, s^m\, F_{mi} \Big(\frac{1}{2}\, \partial_{X j} - \partial_{s j} \Big) + \frac{i}{2}\, s^m F_{mi} \Big(\frac{1}{2}\, \partial_{X i} 
- \partial_{s i} \Big) \Big] \,{\tilde G}_v, \nonumber\\
\eea
it is to be mentioned that the product $\sigma_{\perp i}\, \sigma_{\perp j} = \delta_{ij} - v_i\, v_j - i\, \varepsilon_{jkm}\, v_i\, v_k\, \sigma_m 
- i \varepsilon_{ikm}\, v_j\, v_k\, \sigma_m + i\, \varepsilon_{ijk}\, \sigma_k$. We note that ${\tilde G}_v \propto P_+$ for right handed fermions and 
to obtain the physical interpretations, we have to take the trace by replacing $\sigma_k \rightarrow v_k$. Thus, the product yields the
following: $\sigma_{\perp i}\, \sigma_{\perp j} = \Delta_{ij} + i \,\varepsilon_{ijk}\, v^k$ with $\Delta_{ij} = \delta_{ij} - v_i \,v_j$.
Finally, the sum and difference of EOM at order $O(1/\mu)$ are
\bea \label{ap7}
I^{(1)}_+ &=& \int_s e^{i (l - A) \cdot s} \frac{1}{2\mu} \Big[ \Big(\frac{1}{2} \partial_{X i} \partial_{X j}
+ 2 \partial_{s i} \partial_{s j} + \frac{i}{2}\, s^m F_{mi} \partial_{X j} + \frac{i}{2}\, s^m F_{mj} \partial_{X i} \Big) \Delta_{ij} \nonumber\\
 &-& \varepsilon_{ijk} F_{ij} v_k\Big] {\tilde G}_v 
  = \frac{1}{\mu} \Big[- \bl^2_\perp +  {\bold B} \cdot {\bold v} \Big]\, {\tilde G}_v (X, l), \nonumber\\
I^{(1)}_- &=& \int_s e^{i (l - A) \cdot s} \frac{1}{2\mu}\, \Big[ \Big( \partial_{X i} \partial_{s j}
+  \partial_{X j} \partial_{s i} + i\, s^m F_{mi} \partial_{s j} + i\, s^m F_{mj} \partial_{s i} \Big) \Delta_{ij} ]\, {\tilde G}_v  \nonumber\\
&=& \frac{i}{\mu}\, \bl^\mu_\perp\, \Big( \pd_\m - F_{\m\n}\frac{\pd}{\pd l_\n} \Big)\, {\tilde G}_v (X, l),
\eea
in the above higher order terms $\pd_{Xi} \pd_{Xj}$ and $F_{mi} \pd_{Xj}$ have been ignored. 
Now, the $I^{(2)}_{\pm}$ terms at order $O(1/\mu^2)$ can be simplified in the similar manner to result into the following:
\begin{align} \label{ap8}
I^{(2)}_{+} &= \int_s  e^{i (l - A) \cdot s}\, \frac{1}{4\mu^2}\, \Big[ i\, F_{ij} \pd_{s\mu} + i \,F_{i\mu} \pd_{sj} 
+ i\, F_{\mu j} \pd_{sj} + 2\, \pd_{si}\, \pd_{sj}\, \pd_{s \mu}
\Big]\, \sigma_{\perp i}\, \sigma_{\perp j}\, (-i\, {\bar v}^\mu)\, {\tilde G}_v(x, y), \nonumber\\
I^{(2)}_{-} &= \int_s e^{i (l - A) \cdot s}\, \frac{1}{4\mu^2}\, \Big[ \frac{i}{2}\,  F_{ij}\,  \D_{\mu} + \frac{i}{2} \, F_{i\mu} \D_j +\frac{i}{2}\, F_{\mu j}\, \D_i
\Big] \,\sigma_{\perp i}\, \sigma_{\perp j}\, (-i\, {\bar v}^\mu)\, {\tilde G}_v(x, y),
\end{align}
where we have kept terms to the lowest order in $O(\pd_X)$ and $O(l)$. It should be noted that in the above equation index $\mu = 0, 1, 2, 3$ with ${\bar v}^\mu = (1, - \vv)$. We take the trace over the sigma matrices 
and simplify the above equation which yields  
\bea\label{ap9}
&& I^{(2)}_+ =\frac{1}{4\mu^2}\, \Big[4l_\parallel\, \Big( l^2_\perp - {\bold B} \cdot {\bold v} \Big) + 2 {\bold B} \cdot {\bold l_\perp} 
+ 2({\bold E} \times {\bold l})\cdot {\bold v} \Big] \,{\tilde G}_v (X, l), \nonumber\\
&&I^{(2)}_- = - \frac{i}{4\mu^2} \Big[ \Bigl( 4l_\parallel\, l^\mu - {\bar v}^\mu\, (l^2_\perp - {\bold B} \cdot {\bold v}) \Bigr) 
\D_\m - \Bigl(\varepsilon^{ijk} v^k {\bar v}_\mu F^{i\mu} \Bigr) \D_j  \Big] \,{\tilde G}_v (X, l).
\eea
From the above equation, at order $O(1/\mu^2)$, we can clearly see the difference by a numerical factor in $I^{(2)}_\pm$ 
with SY.

\bibliographystyle{unsrt}
\bibliography{CKT}

\end{document}